\documentclass[prb,twocolumn,showpacs]{revtex4-1}

\usepackage{graphicx}
\usepackage{bm}
\usepackage{verbatim}
\usepackage{amsfonts}
\usepackage{color}
\usepackage{amsmath}

\def\lsim{\mathrel{\rlap{\lower4pt\hbox{\hskip1pt$\sim$}}
    \raise1pt\hbox{$<$}}}                
\def\gsim{\mathrel{\rlap{\lower4pt\hbox{\hskip1pt$\sim$}}
    \raise1pt\hbox{$>$}}}                

\def\Tr{{\text{Tr}}\,}

\def\be{\begin{equation}}
\def\ee{\end{equation}}
\def\bea{\begin{eqnarray}}
\def\eea{\end{eqnarray}}
\def\bse{\begin{subequations}}
\def\ese{\end{subequations}}

\def\Tr{{\text{Tr}}\,}

\def\be{\begin{eqnarray}}
\def\ee{\end{eqnarray}}

\newcommand{\im}{\mathrm{Im}}
\newcommand{\kk}{\mathbf{k}}
\newcommand{\htau}{\hat{\tau}}

\begin{document}

\title{Topological Weyl Superconductor to Diffusive Thermal Hall Metal Crossover\\ in the $\boldsymbol{B}$-Phase of UPt$_3$}
\author{Pallab Goswami}
\affiliation{National High Magnetic Field Laboratory and Florida State University, Tallahassee, Florida 32310, USA}
\altaffiliation{Present address: Condensed Matter Theory Center and Joint Quantum Institute, Department of Physics, University of Maryland, College Park, Maryland 20742-4111, USA.}
\author{Andriy H. Nevidomskyy}
\email{nevidomskyy@rice.edu}
\affiliation{Department of Physics and Astronomy, Rice University, TX 77005, USA}

\begin{abstract}
The recent phase sensitive measurements in the superconducting $B$-phase of UPt$_3$ provide strong evidence for the triplet, chiral $k_z(k_x \pm ik_y)^2$ pairing symmetries, which endow the Cooper pairs with orbital angular momentum projections $L _z= \pm 2$ along the $c$-axis. In the absence of disorder such pairing can support both line and point nodes, and both types of nodal quasiparticles exhibit nontrivial topology in the momentum space. The point nodes, located at the intersections of the closed Fermi surfaces with the $c$-axis, act as the double monopoles and the antimonopoles of the Berry curvature, and generalize the notion of Weyl quasiparticles. Consequently, the $B$ phase should support an anomalous thermal Hall effect, the polar Kerr effect, in addition to the protected Fermi arcs on the (1,0,0) and the (0,1,0) surfaces. The line node at the Fermi surface equator acts as a vortex loop in the momentum space and gives rise to the zero energy, dispersionless Andreev bound states on the (0,0,1) surface.  At the transition from the $B$-phase to the $A$-phase, the time reversal symmetry is restored, and only the line node survives inside the $A$-phase. As both line and double-Weyl point nodes possess linearly vanishing density of states, we show that weak disorder acts as a marginally relevant perturbation. Consequently, an infinitesimal amount of disorder destroys the ballistic quasiparticle pole, while giving rise to a diffusive phase with a finite density of states at the zero energy. The resulting diffusive phase exhibits $T$-linear specific heat, and an anomalous thermal Hall effect. We predict that the low temperature thermodynamic and transport properties display a crossover between a ballistic thermal Hall semimetal and a diffusive thermal Hall metal. By contrast, the diffusive phase obtained from a time-reversal invariant pairing exhibits only the $T$ linear specific heat without any anomalous thermal Hall effect.
\end{abstract}

\pacs{03.65.Vf, 
74.20.-z, 
74.20.Rp, 
74.70.Tx 
}


\maketitle

\section{Introduction} Topological states of matter are usually characterized as fully gapped insulating or superconducting states in the bulk with gapless excitations on the surface. Recently it has been recognized that gapless systems can also possess nontrivial momentum space topology and protected surface states~\cite{Wan,Volovik,Beri, Schnyder, Kopnin}. Being gapless in the bulk, these states fall outside the classification scheme of the gapped topological insulating and superconducting states~\cite{Schnyder_2008,Kitaev_2009,Ryu_2010}. Some interesting examples of such gapless systems in three dimensions are semimetallic phases, where two non-degenerate energy bands touch either at isolated point nodes or Weyl points~\cite{Wan,Volovik}, or along a line node~\cite{Beri,Schnyder, Kopnin}. These systems constitute intriguing examples of fermionic quantum critical systems~\cite{GoswamiChakravarty} (power law behaviors of thermodynamic quantities) with nontrivial momentum space topology.

The quasiparticles possess linear dispersion in the vicinity of the Weyl points, and in particular the Weyl points act as the monopoles and anti-monopoles of unit strength of the Berry curvature, and the topological invariant for these systems is determined by the strength of the (anti)monopoles ($\pm 1$), which also determines the chirality or the handedness of the quasiparticles. In the presence of translational invariance, Weyl fermions are topologically protected, in the sense that they can only be eliminated when a pair of such point nodes with opposite handedness (chirality) merge in momentum space due to  the tuning of some external parameter \cite{Volovik}. A number of systems have been recently shown~\cite{Wan,Volovik,Burkov1,Burkov2,Zyuzin1,Cho,Liu,Gong,Sau,Meng,Das,GoswamiTewari1,GoswamiBalicas} to support massless Weyl fermions (in either semi-metal or superconducting phases) as the bulk low energy excitations, and the associated topologically protected zero-energy surface states in the form of open Fermi/Majorana arcs. There are also interesting proposals for systems which can exhibit line nodes in the bulk and dispersionless zero-energy bound states on the surface~\cite{Schnyder,Kopnin,GoswamiBalicas}.

However, these gapless systems are yet to be experimentally realized, the only exception being the Weyl fermions around the Fermi surface poles in the $A$-phase of $^3$He~[\onlinecite{Volovik}]. Therefore, it behooves us to identify promising experimental systems for realizing gapless topological states of matter. In this direction, gapless superconducting systems are expected to play an important role. In particular the charge conjugation symmetry of superconductor guarantees that the chemical potential of the nodal excitations is exactly tuned at the band touching points or lines. In a recent paper~[\onlinecite{GoswamiBalicas}], one of us has argued that the uranium based unconventional superconductors with both line and point nodes can serve as multi-functional systems. Motivated by this, we critically examine the nodal topology of the superconducting $B$ phase of the heavy-fermion compound UPt$_3$. Our main findings are summarized below:

\begin{enumerate}

\item We show that the superconducting $B$ phase of the heavy-fermion compound UPt$_3$ provides a spectacular example of triplet, chiral $f$-wave pairing, which supports a line node in the basal $ab$ plane and double Berry (anti)monopoles at the intersection of the closed Fermi surfaces with the $c$-axis. We demonstrate that the line node acts as a vortex loop in the momentum space, while the double Berry monopole (of strength $\pm2$) generalizes the notion of the Weyl fermions (unit monopoles) to double Majorana--Weyl fermions.

\item The line node is expected to give rise to dispersionless, zero-energy surface Andreev bound states (SABS) on the $(0,0,1)$ surface. On the other hand, the double monopoles give rise to two protected Fermi arcs on the $(1,0,0)$ and $(0,1,0)$ surfaces, which are bounded by the images of the monopole and the antimonopole on the surface Brillouin zone. \emph{In the absence of disorder, the ballistic double Majorana--Weyl fermions give rise to a large anomalous thermal Hall conductivity $\kappa^c_{xy} \sim 10^{-3} W K^{-1} m^{-1}$, which is measurable with conventional experimental set up}. We also predict a topological phase transition between the $A$ and the $B$ phase.

\item Finally, we consider the effects of weak disorder on the nodal quasiparticles. Both types of nodal quasiparticles possess linearly vanishing density of states $D(\epsilon) \propto \epsilon$. Consequently, weak disorder acts as a marginally relevant perturbation that gives rise to a finite density of states at zero energy. In this process, weak disorder vanquishes the ballistic quasiparticle poles. \emph{Therefore, the line node and the double-Weyl point nodes become unstable against an infinitesimal amount of disorder, which converts the topological semimetallic phase of the BCS quasiparticles into a diffusive phase}. Immediately below the transition temperature, the specific hxeat varies as $T^2$ (due to the linear density of states associated with ballistic excitations) and eventually crosses over to $C_v \sim T$ (characteristic of a diffusive phase). Due to the broken time reversal symmetry, the diffusive phase still supports anomalous thermal Hall effect. We argue that the value of $\kappa_{xy}$ in this diffusive metal is comparable to that in the ballistic double Majorana--Weyl regime, for smooth disorder that does not cause appreciable backscattering between two nodes of opposite chirality. By contrast, the diffusive phase emerging from of a time-reversal invariant $A$-phase with line nodes does not support anomalous thermal Hall effect, even though it shows the crossover behavior for the specific heat similar to the $B$-phase.

\end{enumerate}

The manuscript is organized as follows. In Sec.~\ref{phenomenology}, we review the pertinent experimental results for the $B$ phase. In Sec.~\ref{invariant} we describe the bulk topological invariants associated with the nodal quasiparticles. In Sec.~\ref{surfacestates} we consider the bulk--boundary correspondence and the two different types of surface Andreev bound states localized on $(0,0,1)$ and $(1,0,0)$ surfaces, which are  related to the topological invariants of the line and point nodes, respectively. In Sec.~\ref{thermalHall} we show how the Berry curvature of the  double Majorana--Weyl quasiparticles gives rise to a large anomalous thermal Hall conductivity, and provide an order of magnitude estimate of $\kappa_{xy}$. The summary of the main results and a discussion of the relevant effects of weak quenched disorder on the nodal excitations are provided in Sec.~\ref{conclusion}. The explicit calculations of the surface Andreev bound states on $(0,0,1)$ and $(1,0,0)$ surfaces are  presented in Appendix~\ref{Appendix1} and Appendix~\ref{Appendix2} respectively.

\section{Phenomenology of the $\boldsymbol{B}$-phase}\label{phenomenology}The heavy fermion compound UPt$_3$ is one of the best studied and most convincing candidates for the nodal superconductivity. The power-law temperature dependence of the ultrasonic attenuation~\cite{Bishop84},
the NMR relaxation rate ($T^{-1}_1 \sim T^3$)~\cite{Kohori88},  magnetic field penetration depth~\cite{Broholm90} and thermal conductivity~\cite{Suderow97} suggest the existence of nodal excitations with density of states (DOS) $D(\epsilon) \propto \epsilon$. The linear in energy DOS is also supported by the magnetic field dependence of the specific heat $C\propto \sqrt{H}$~\cite{Ramirez95}.
It has been realized early on
that the superconducting order parameter satisfying this requirement must belong to either the pseudospin-singlet $E_{2g}$~\cite{Joynt92} or the pseudospin-triplet $E_{2u}$~\cite{Norman92, Sauls94} irreducible representation of the hexagonal $D_{6h}$ point group. Early attempts to unambiguously discriminate between these two possibilities based on existing measurements proved inconclusive~\cite{Norman96, Graf00, Joynt02}. If however one takes into account the evidence from NMR Knight shift measurements~\cite{Tou96,Tou98} in favor of spin-triplet pairing, the natural choice is the $E_{2u}$ triplet order parameter. Recently, this conjecture has received strong support from the phase-sensitive Josephson interferometry~\cite{Strand09, Strand10,footnote1}.

The $E_{2u}$ pseudospin-triplet order parameter is characterized by two basis functions, $\Delta_{1,\mathbf{k}} \sim k_z(k_x^2-k_y^2)$ and $\Delta_{2,\mathbf{k}} \sim 2k_z k_x k_y$ multiplying the  $\mathbf{d}$-vector $\mathbf{d} \propto \hat{z}$. The resulting $f$-wave order parameter is then given by $\Delta_{\alpha,\beta}(\mathbf{k}) = \Delta_0(\sigma_z\cdot i\sigma_2)[\eta_1 \Delta_{1,\mathbf{k}} + i\eta_2 \Delta_{2,\mathbf{k}}]$, which is similar to the polar phase of $^3$He, but with the additional $(k_x,k_y)$-dependence of the pairing wavefunctions encoded in $\Delta_{1/2,\mathbf{k}}$.
The choice of the coefficients $\eta_1$ and $\eta_2$ in the above expression for $\Delta(\mathbf{k})$ is determined by energetic considerations, and in zero external magnetic field, two different phases are realized: the $A$-phase with $\eta_2=0$ is stable at high temperatures  ($T_c^B\!<\!T\!<\!T_c=550$~mK) and the low-temperature $B$-phase with $\eta_2= \eta_1$ is realized at $T<T_c^B=500$~mK. The latter choice results in the pairing wavefunction $\Delta_B(\mathbf{k}) \sim k_z(k_x+i k_y)^2$ which is time reversal symmetry breaking, chiral pairing and has U(1) axial symmetry.
The pairing wavefunction can be expressed in terms of spherical harmonics.
We find that $\Delta_A(\kk)\sim (Y_3^{+2} + Y_3^{-2})$, i.e. the $A$-phase has zero net orbital momentum, $L_z\!=\!0$. By contrast, the $B$-phase, $\Delta_B(\kk)\!\sim Y_2^{+2}\cdot Y_1^{0} \propto Y_3^{+2}$ is characterized by non-zero orbital momentum \mbox{$L_z\!=\!+2$}.

The nature of chiral pairing in the $B$ phase has been confirmed by the Josephson interferometry in a corner-junction setup~\cite{Strand09}, demonstrating a phase shift $\pi$ of the pairing wavefunction upon the $90^\circ$ rotation about the $c$-axis. Both the $A$ and $B$ phase possess gapless line nodes at the equator $k_z=0$, consistent with NMR and thermal conductivity data~\cite{Kohori88,Suderow97}.  However only the low-temperature $B$ phase has point nodes at the intersections of the Fermi surfaces with the $c$ axis. The $A$-phase, on the other hand, has additional line nodes at $k_x=\pm k_y$~\cite{Strand10}.

The $E_{2u}$ gap symmetry deduced from phase-sensitive measurements~\cite{Strand09}  has been challenged in Refs.~[\onlinecite{Machida12,Tsutsumi1,Tsutsumi2}], based on the field angle-dependent thermal conductivity measurements,  instead proposing a planar $E_{1u}$ gap symmetry: $\mathbf{d}_{1u}\sim(\hat{\mathbf{y}} k_x + \hat{\mathbf{z}} k_y) (5k_z^2 - 1)$.
We argue that the two scenarios can be distinguished based on the time reversal symmetry (TRS) breaking signature: the $E_{2u}$ phase obviously breaks TRS whereas the planar $E_{1u}$ phase does not.
Experimentally, TRS in $B$ phase was first indicated~\cite{Luke93} by the muon spin resonance ($\mu$SR) measurements\cite{footnote2}.
Very recently, the same conclusion was reached by the observation of a field-trainable polar Kerr effect, which is only present in the $B$ phase but disappears in the $A$ phase~\cite{Kapitulnik}. It has also been confirmed that the polar Kerr effect is not related to the presence of a small antiferromagnetic moment in the basal $ab$ plane. Together, Josephson interferometry and the polar Kerr effect measurements provide compelling evidence in favor of the chiral $E_{2u}$ state~\cite{footnote3}. We note that very recently, after this manuscript was submitted, an alternative ``chiral $E_{1u}$" scenario has been proposed\cite{Izawa2014}, characterized by the $d$-vector $\mathbf{d}_{1u}^{chiral}\sim \hat{\mathbf{z}} (k_x + ik_y) (5k_z^2 - 1)$. Such a chiral $E_{1u}$ order parameter would be consistent with the observed TRS breaking, although it contradicts the Josephson interferometry results\cite{Strand09}.


\section{Topological invariants in the bulk} \label{invariant}After defining the Nambu spinor $\Psi^\dagger_{\mathbf{k}}=(c^\ast_{\mathbf{k},\uparrow}, c_{-\mathbf{k},\downarrow}, c^\ast_{\mathbf{k},\downarrow}, c_{-\mathbf{k},\uparrow})$, the reduced BCS Hamiltonian for $S_z=0$ triplet pairing acquires the following block-diagonal form
\begin{eqnarray}
\hat{H}_{\mathbf{k}}=\hat{h}_{\mathbf{k}}\otimes \sigma_0=\mathbf{N}_{\mathbf{k}}\cdot \boldsymbol{\hat{\tau}} \otimes \sigma_0, \label{eqn:hk}
\end{eqnarray}
where $\mathbf{N}_{\mathbf{k}}=(\mathrm{Re}(d_z),\mathrm{Im}(d_z),\xi_{\mathbf{k}})$, $\hat{\tau}_i$ are Pauli matrices acting in Nambu space, and  $\sigma_0$ is the unity matrix in the spin space. For a simplified discussion of the topological properties of the nodal excitations, we assume a parabolic dispersion $\xi_{\mathbf{k}}=\mathbf{k}^2/(2M)-\mu$, with $\mathrm{Re}(d_z)=\Delta_0/(k^3_F)k_z(k^2_x-k^2_y)$, and $\mathrm{Im}(d_z)=2\Delta_0/(k^3_F)k_xk_yk_z$, without the loss of generality. This chiral $f$-wave pairing supports a line node along the Fermi surface equator in the $ab$ plane described by $\mathbf{k}=(k_F \cos \phi_k, k_F\sin \phi_k,0)$, where $0 \leq \phi_k\!=\!\arctan(k_y/k_x) \leq 2\pi$ is the azimuthal angle. In addition, there are two point nodes at the poles of the Fermi surface intersecting the $c$-axis: $\mathbf{k}=(0,0,\pm k_F)$. Both types of nodal excitations can be classified in terms of appropriate topological invariants in the following way.

\begin{figure*}[tb]
\includegraphics[width=0.8\textwidth]{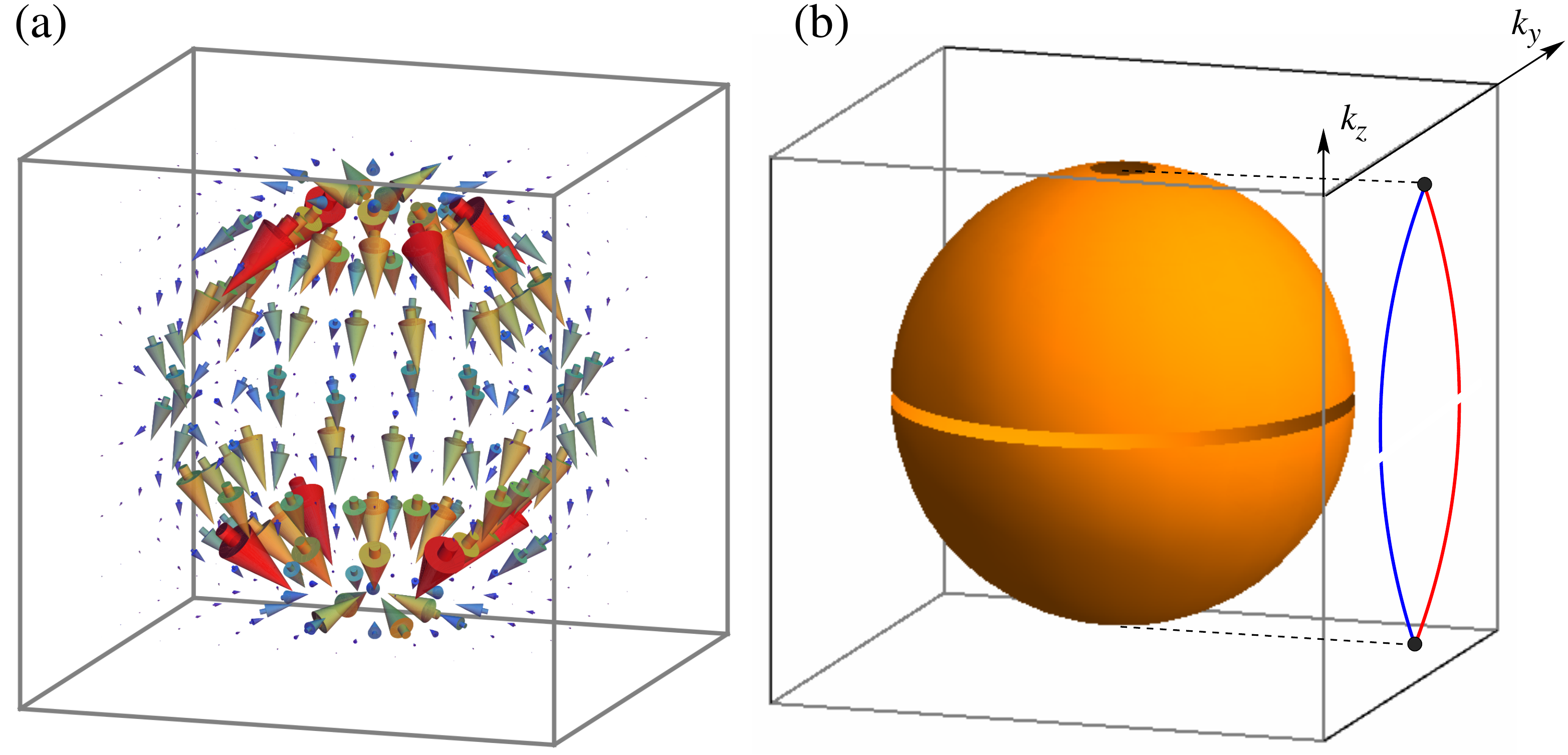}
\caption{(Color online) (a) A vector-field plot of the quasiparticle's Berry curvature $\Omega_{\kk,z}$ for $\Delta_{\mathbf{k}}= \Delta_{0}k_z(k_x - ik_y)^2/k^3_F$. The point nodes at $k_z=\pm k_F$ appear as the double (anti)monopole of the Berry curvature. The Berry curvature is diminished at the equator $k_z=0$, due to the presence of the line node. (b) Sketch of the (spherical) Fermi surface with projection onto the (100)-surface Brillouin zone, showing the double-Weyl points at $k_z=\pm k_F$ connected by two Andreev surface arcs.}
\label{fig:1}
\end{figure*}

The Bloch wavefunction of the BCS quasiparticle changes sign when encircling the line node along a closed loop, and consequently the line node acts as a vortex loop in the momentum space~\cite{Beri, Schnyder,Kopnin}. In order to describe the topological invariant for the nodal ring, we first note that the BCS Hamiltonian possesses a spectral symmetry with respect to the unitary transformation $\mathcal{U}_{\mathbf{k}}=(-\sin 2\phi_{\mathbf{k}} \tau_1+\cos 2\phi_{\mathbf{k}} \tau_2)$, i.e., $ \{ \mathcal{U}_k, \hat{h}_k \}=0$.  Due to this symmetry, if $\psi_k$ is a wavefunction of energy $E$, $\mathcal{U}_k \psi_k$ becomes the wavefunction with energy $-E$. With the help of this unitary matrix, we can define the following topological invariant for the line node
\begin{eqnarray}\label{tinvring}
N_{\mathcal{L}}=-\frac{1}{4\pi i} \oint_\mathcal{L}  \mathrm{d}l \; \Tr\!\left[\mathcal{U}_{\mathbf{k}} h^{-1}_{\mathbf{k}} \partial_{l} h_{\mathbf{k}}\right],
\end{eqnarray}
where $\mathrm{d}l$ is the line element along a closed loop $\mathcal{L}$ encircling the nodal line. After an explicit calculation following Ref.~\onlinecite{GoswamiBalicas} we obtain $N_{\mathcal{L}}=1$. However, we note that the topological invariant of the line node does not depend on the angular momentum $L_z$ of the Cooper pair. For this reason, even when the pairing amplitude is real ($\im(d_z)=0$) in the $A$ phase, we can adopt a similar definition for the $N_{\mathcal{L}}$, after identifying the spectral symmetry matrix to $\tau_2$. Due to this nontrivial topological property of the line node, we expect the existence of zero-energy surface Andreev bound states on the $(0,0,1)$ surface in both the $A$ and $B$ phases of UPt$_3$.

By contrast, the topological invariant of the point nodes depends on $L_z$. For the Bloch wavefunctions $\psi_{\mathbf{k},n}$ of $\hat{h}_{\mathbf{k}}$, where $n =\pm 1$ is the band index, there exists a fictitious vector potential $\mathcal{A}_{\mathbf{k},n}=i \langle \psi_{\mathbf{k},n} | \nabla_{\mathbf{k}} | \psi_{\mathbf{k},n} \rangle$. The gauge-invariant field strengths or the Berry curvatures are defined as
\begin{equation}\label{Berry}
\Omega_{\mathbf{k},n,a}=\nabla_{\mathbf{k}} \times \mathcal{A}_{\mathbf{k},n}=\frac{(-1)^n}{4} \; \epsilon_{abc} \; \mathbf{n}_{\mathbf{k}} \cdot \bigg[ \frac{\partial \mathbf{n}_{\mathbf{k}}}{\partial k_b}\times \frac{\partial \mathbf{n}_{\mathbf{k}}}{\partial k_c}\bigg].
\end{equation}
For $L_z=2$ state, the explicit expressions for the three components of $\Omega_{\mathbf{k},n,a}$ are given by
\begin{eqnarray}
\Omega_{\mathbf{k},n,x}\!\!&=&\!\! \frac{(-1)^{n+1}\mathrm{sgn}(L_z) \; \Delta^2_0\;k_zk_xk^2_\perp(k^2_z-k^2_\perp+k^2_F)}{\mu^2 k^2_F \; \left[(k^2-k^2_F)^2+\frac{\Delta^2_0}{\mu^2k^2_F}k^2_zk^4_\perp\right]^{\frac{3}{2}}} \phantom{\hspace{4mm}} \\
\vspace{2mm} \nonumber\\
\Omega_{\mathbf{k},n,y}\!\!&=&\!\! \frac{(-1)^{n+1}\mathrm{sgn}(L_z) \; \Delta^2_0\;k_zk_yk^2_\perp(k^2_z-k^2_\perp+k^2_F)}{\mu^2 k^2_F \; \left[(k^2-k^2_F)^2+\frac{\Delta^2_0}{\mu^2k^2_F}k^2_zk^4_\perp\right]^{\frac{3}{2}}} \phantom{\hspace{4mm}} \\
\vspace{2mm} \nonumber\\
\Omega_{\mathbf{k},n,z}\!\!&=&\!\! \frac{(-1)^{n+1}\mathrm{sgn}(L_z) \; 2\Delta^2_0\;k^2_zk^2_\perp(k^2_z-k^2_F)}{\mu^2 k^2_F \; \left[(k^2-k^2_F)^2+\frac{\Delta^2_0}{\mu^2k^2_F}k^2_zk^4_\perp\right]^{\frac{3}{2}}}.\ \phantom{\hspace{4mm}} \label{eq:Omega_z}
\end{eqnarray}
Notice that $\Omega_{\mathbf{k},n,x}$ and $\Omega_{\mathbf{k},n,y}$ are odd functions of their arguments. As a result, the number of field lines coming in and out of the $ca$ or the $cb$ planes are equal, and there is no net flux through the $ca$ and the $cb$ planes. In contrast, $\Omega_{\mathbf{k},n,z}$ in Eq.~(\ref{eq:Omega_z}) is an even function of its arguments, and the flux through the $ab$ plane as a function of $k_z$ equals
\begin{equation}
\Phi_\Omega(k_z)= \int d^2 k_\perp \Omega_{\mathbf{k},n,z}=2\pi\, \mathcal{C}(k_z)
\end{equation}
where $\mathcal{C}(k_z)$ is the Chern number for the effective two-dimensional problem for a fixed $k_z$. For a given value of  $-k_F\!\! <\!\! k_z\!\!<\!\! k_F$ (provided $k_z\!\neq\! 0$),
$\hat{h}_{\mathbf{k}}$ describes an effective two-dimensional problem with fully gapped weak/BCS pairing and an effective chemical potential $\tilde{\mu}=\mu-k^2_z/(2M)$, which leads to $\mathcal{C}(k_z)=L_z$. For $k^2_z> k^2_F$, the effective chemical potential of the two-dimensional problem becomes negative, and describes topologically trivial BEC pairing, as $\mathcal{C}(k_z)$ vanishes. Therefore, point nodes $k_z=\pm k_F$ act as the monopoles and antimonopoles of the Berry curvature of strength $L_z$, and the flux through a sphere surrounding the (anti)monopole is equal to $\pm L_z$. Thus the topological invariant of the nodal points is precisely determined by the Cooper pairs's angular momentum projection $L_z$ along the $c$-axis. We have plotted the Berry curvature for $L_z=-2$ pairing and $n= +1$ band in Fig.~\ref{fig:1}a, which captures all the salient features described above. For $L_z=+2$, the monopole and the antimonopole swap places. By expanding $\hat{h}_{\mathbf{k}}$ for $L_z=-2$ around the point nodes we obtain the low energy Hamiltonians
\begin{equation}\label{doubleWeyl}
\hat{h}_{R/L}=\pm v_F(k_z \mp k_F)\htau_3 \pm \frac{\Delta_0}{k^2_F}\left((k^2_x-k^2_y) \htau_1+2k_x k_y \htau_2 \right),
\end{equation}
for the double (anti)monopole. The low energy excitations around the point nodes possess linear  dispersion along the $c$ axis and quadratic dispersion in the $ab$ plane. This anisotropic dispersion leads to the linear density of states $D(\epsilon) \propto \epsilon$. The nontrivial Chern number for $-k_F \!< \!k_z\!< \!k_F$ (and $k_z\! \neq\! 0$) is responsible for giving rise to the current carrying, chiral Andreev bound states on the $(1,0,0)$ and the $(0,1,0)$ surfaces,
and many exotic chiral transport properties to be discussed below.

\section{Surface Andreev bound states} \label{surfacestates}As mentioned above, the bulk invariant $L_z=2$ for the point nodes has already been determined by the Josephson interferometry measurements~\cite{Strand09}. However surface-sensitive techniques, such as ARPES and Fourier transformed STM, can provide more direct information about the topological surface states, tied to the existence of the two types of bulk topological invariants. For this reason, we now consider the different types of surface Andreev bound states arising from the line and the point nodes.

We first describe the zero-energy bound states due to the existence of the line node on the $(0,0,1)$ surface. We consider a semi-infinite sample, such that $z<0$ and $z>0$ regions are  occupied by the superconductor and the vacuum, respectively. Since $k_x$ and $k_y$ are good quantum numbers, the zero-energy bound states satisfy the differential equations $\hat{h}(k_x,k_y,-i\partial_z) \psi(\kk_\perp, z)=0$, and the boundary conditions $\psi(z=0)=\psi(z \to -\infty)=0$. We have already noted that the presence of a spectral symmetry with respect to the matrix $\mathcal{U}_{\mathbf{k}}$ (independent of $k_z$), which continues to be valid even after the replacement $k_z \to -i \partial_z$. For this reason, the zero-energy bound states need to be eigenstates of $\mathcal{U}_{\mathbf{k}}$ with eigenvalue $+1$ or $-1$. Actually, there is an index theorem~\cite{Volovik,Tanaka1,Tanaka2}
\begin{equation}\label{index1}
\langle \psi| \mathcal{U}_\mathbf{k} |\psi \rangle=n_+-n_-=N_{\mathcal{L}}(z=-\infty)-N_{\mathcal{L}}(z=+\infty),
\end{equation}
where $n_\pm$ respectively represent the number of zero energy states with eigenvalues $\pm 1$. This index theorem guarantees the existence of the zero energy surface states between the topologically trivial and non-trivial regions and succinctly describes the bulk-boundary correspondence. An explicit calculation detailed in Appendix~\ref{Appendix1}, shows that $n_+=1$ and $n_-=0$ for the normalizable zero-energy bound state, for any $k_\perp < k_F$. Therefore, the zero-energy surface states produce an image of the Fermi surface equator in the $ab$ plane. Being dispersionless, these zero-energy states result in a divergent density of states, and are predicted to give rise to a zero bias peak in tunneling measurements. These zero modes are Majorana fermions, of which there are two copies, arising from the twofold spin degeneracy of the pairing interaction in Eq.~(\ref{eqn:hk}). Any perturbation mixing the two spins will lead to the hybridization of these two Majorana modes.
It is also important to note that a Zeeman interaction always gaps out these surface states. Therefore, a comparative study of the STM measurements on the $(1,0,0)$ surface with and without a magnetic field may provide direct evidence of the line node in the $ab$ plane. In the presence of spin-orbit coupling, these zero energy surface states can also be eliminated by the surface Rashba coupling~\cite{Kobayashi}.

The presence of the nontrivial Chern number $\mathcal{C}(k_z)=L_z=\pm2$ for $|k_z|< k_F$ (except at the nodes), guarantees the existence of \emph{two} chiral edge modes (often referred to as ``Fermi arcs'') for the effective two dimensional problem, obtained by fixing $k_z$. Generally, we predict the number of the chiral surface Fermi arcs to be equal to $L_z$, generalizing the case of Weyl fermions in Ref.~\onlinecite{Wan} to $L_z>1$.
In the surface Brillouin zones of $(1,0,0)$ and $(0,1,0)$ surfaces, these chiral Andreev bound states only exist for $|k_z|<k_F$ and $k_z \neq 0$. After an explicit calculation for the SABS on the $(1,0,0)$ surface for $L_z=-2$ pairing, as detailed in Appendix~\ref{Appendix2}, we obtain two Fermi arcs with
$\sqrt{2}k_y=\pm \sqrt{k^2_F-k^2_z}$ plotted schematically in Fig.~\ref{fig:1}b.
These SABS are two-fold degenerate, chirally dispersing (along the $y$ direction) Majorana fermions, which can be directly probed by  Fourier transformed STM measurements. The number of Fermi arcs determined from STM also provides a check  of the underlying angular momentum $L_z$.
These chiral surface states also carry current, which in principle can be measured by SQUID microscopy\cite{Moler1}.

We note in passing that the recently proposed alternative gap symmetry scenario, the so-called \mbox{``chiral $E_{1u}$"~[\onlinecite{Izawa2014}]}, is characterized by the $d$-vector $\mathbf{d}_{1u}^{chiral}\sim \hat{\mathbf{z}} (k_x + ik_y) (5k_z^2 - 1)$, so it has $L_z=\pm 1$. If indeed realized in the $B$-phase of UPt$_3$, it would have a monopole/antimonopole of unit strength at the north/south pole of the Fermi surface and would fall into the category of ``Weyl superconductors". Consequently, it would have \emph{one} chiral Fermi arc on the surface, as opposed to two considered above. Therefore, if it were possible to experimentally measure the number of the chiral Andreev surface states in the $B$-phase of UPt$_3$, it would provide a conclusive resolution to the ongoing debate about the $k_{xy}$-dependence of the superconducting order parameter.

\section{Anomalous Thermal Hall conductivity} \label{thermalHall}The presence of Berry curvature and the chiral surface states can lead to many interesting chiral transport properties and magnetoelectric effects for the $B$ phase~\cite{GoswamiBalicas,GoswamiTewari2,Lutchyn,Kallin}. Here we only consider the existence of a large anomalous thermal Hall effect. The thermal Hall conductivity of the chiral quasiparticles is determined by
\begin{equation}\label{tHall}
\kappa_{ab}=-\frac{k^2_B T}{\hbar} \epsilon_{abc}\int \frac{d^3 k}{(2\pi)^3} \; \Omega_{n,\mathbf{k},c} \; \int_{E_{n,\mathbf{k}}}^{\infty} dE \; E^2 \; \frac{\partial f(E)}{\partial E},
\end{equation}
which in the low temperature limit leads to $\kappa^0_{xy}= \frac{\pi^2 k^2_B T}{3h} \times L_z \times \left(\frac{\Delta k}{2\pi}\right)$, where $\Delta k = 2k_F$ is the distance between the double-Weyl points in momentum space. Since $\kappa_{xy}$ vanishes at $T=0$ and $T^B_c$ (due to restoration of time reversal symmetry in the $A$ phase), the maximum value of $\kappa_{xy}$ should occur at a model dependent temperature $0<T<T^B_c$. An explicit calculation using Eq.~(\ref{tHall}) shows this temperature to be roughly $T^B_c/2$. The maximum value of $\kappa_{xy}$ is of order
\begin{equation}
\kappa^c_{xy}=\frac{\pi^2 k^2_B T^B_c}{3h} \times L_z \times \left(\frac{\Delta k}{2\pi}\right).
\label{eq:kappa}
\end{equation}
As this equation shows, the thermal Hall effect is not quantized, as expected in a metallic state, since it depends on a non-universal distance $\Delta k$ between the double-Weyl points. Nevertheless, the origin of this effect is clearly topological, as $\kappa^c_{xy}$ is proportional to the Chern number $C(k_z) = L_z$. This may provide an additional route towards verifying the symmetry of the superconducting order parameter in UPt$_3$. Indeed, the $E_{2u}$ symmetry deduced from the phase-sensitive measurements~\cite{Strand09} requires $L_z=2$, whereas the chiral $E_{1u}$ symmetry mentioned earlier has $L_z=1$~[\onlinecite{Izawa2014}].

It is instructive to estimate the magnitude of $\kappa_{xy}$ from Eq.~(\ref{eq:kappa}). Substituting $k_F \sim 10^9 m^{-1}$ from the de Haas van Alphen measurements~\cite{McMullan} and $T^B_c \sim 500$~mK, we find $\kappa^c_{xy} \sim 10^{-3} W K^{-1} m^{-1}$, which is likely a lower bound on $\kappa_{xy}$, as it can only increase after accounting for the multiple Fermi surface sheets (five for UPt$_3$). This is a large value for the intrinsic anomalous thermal Hall conductivity and can be easily measured with present experimental accuracy~\cite{Checkelsky12}. Experimental verification of anomalous $\kappa_{xy}$ in the $B$ phase will provide a smoking gun signature of the underlying chiral pairing.

\section{Conclusions}\label{conclusion}
We have so far considered the exotic nodal topology of the chiral $f$-wave pairing $k_z(k_x \pm i k_y)^2$. We have clearly demonstrated that the line node and the double-Weyl point nodes lead to similar thermodynamic properties, as the density of states for both types of quasiparticles is linear $D(\epsilon) \sim |\epsilon|$. We have identified how the bulk invariants lead to different types of protected zero-energy Andreev bound states on $(0,0,1)$ and $(1,0,0)$ surfaces. Our simple estimation of the anomalous thermal Hall conductivity shows that the double Majorana--Weyl fermions possess a large $\kappa_{xy}$.

Finally, we discuss the effects of quenched disorder on the stability of the nodal excitations. We will only consider the random variation of the underlying Fermi level of the normal state quasiparticles. If we calculate the retarded self energy $\Sigma^R(\epsilon)$ by employing self-consistent Born approximation, the imaginary part of the self-energy or the scattering rate $\tau^{-1}$ at zero energy satisfies
\begin{equation}
W \int d\epsilon \; \frac{\rho(\epsilon)}{\frac{\hbar^2}{\tau^2}+\epsilon^2}=1,
\end{equation} where $W$ is the disorder coupling constant.
After substituting the linear density of states, $\rho(\epsilon)=\mathcal{A} |\epsilon| \Theta(E_c-|\epsilon|)$, where $E_c$ is the ultraviolet cutoff and $\mathcal{A}$ is a material-dependent constant, we obtain
\begin{equation}
\frac{\hbar}{\tau}=E_c \; \exp \left(-\frac{1}{W \mathcal{A}}\right).
\end{equation} Therefore, weak disorder acts as a marginally relevant perturbation. As a result, an infinitesimal amount of disorder gives rise to a finite density of states $\sim \tau^{-1}$ at zero energy, and destroys the ballistic quasiparticle pole at zero energy. Consequently, infinitesimally weak disorder destroys the topological properties of the line node and the double-Weyl points for a thermodynamically large system size. When $|\epsilon| > \hbar/\tau$, the quasiparticles exhibit ballistic behavior, and cross over to the diffusive behavior below the energy scale $\hbar/\tau$. By controlling the amount of impurities, it is possible to substantially lower $\hbar/\tau$, which is crucial for observing the exotic properties of the ballistic nodal excitations in a real material.

A finite density of states at zero energy gives rise to a $T$-linear specific heat for the diffusive phase. Hence, the specific heat shows a crossover from the ballistic $\sim T^2$ to a diffusive $\sim T$ linear behavior below $T \sim \hbar/(k_B \tau)$. This holds even for a time-reversal invariant state with only line nodes, such as in the $A$-phase of UPt$_3$ (for earlier work on such crossover effects due to impurities in heavy fermion superconductors see Refs.~\onlinecite{Hirschfeld1,Ott}). However, due to the lack of time reversal symmetry in the $B$-phase, the diffusive phase associated with the chiral pairing also supports an anomalous Hall effect. If the underlying disorder potential has a smooth spatial variation, the backscattering effects between two double-Weyl points can be extremely small, and our estimate for $\kappa_{xy}$ will remain valid. Only for short range disorder, the backscattering effects are very strong and our estimations will not be quantitatively reliable. \\

\section*{Acknowledgements}
We would like to thank Jim Sauls, Yoshiteru Maeno and Yuji Matsuda for fruitful discussions.
P.~G. was supported by the NSF Cooperative Agreement No.~DMR-0654118, the State of Florida, and the U. S. Department of Energy. A.~H.~N. was supported by the Welch Foundation grant C-1818 and the CAREER Award from the National Science Foundation (Grant No. DMR-1350237). 

\appendix

\section{Zero energy Andreev bound state on the $\boldsymbol{(0,0,1)}$ surface} \label{Appendix1}

We are considering a boundary at $z=0$, such that $z<0$ and $z>0$ regions are respectively occupied by the superconductor and the vacuum. The two component spinor wave function $\psi^T=(u,v)$ (for $h(k_x,k_y,-i\partial_z)$) for zero energy surface Andreev bound state satisfies the following differential equations
\begin{eqnarray}\label{surface1}
  \bigg[\big(-\partial^2_z+k^2_{\perp}-k^2_F   \Big)  \tau_3  \phantom{\hspace{5cm}}&\\
 - \frac{i\Delta_0}{\mu k_F}\; k^2_\perp \left(\cos 2\phi_\mathbf{k} \tau_1 + \sin 2\phi_{\mathbf{k}} \tau_2\right)\partial_z\bigg]\psi(k_x,k_y,z)=0,& \nonumber
\end{eqnarray}
and the boundary conditions $\psi(z=0)=\psi(z\to -\infty)=0$. Since, the zero energy wavefunctions are eigenstates of $\mathcal{U}_{\mathbf{k}}$, we substitute
\begin{equation}\label{zeromode1}
\psi_\pm(x,y,z)=e^{i(k_xx+k_yy)} \;F_\pm(z) \; \left(\begin{array}{c}
e^{-\frac{i}{2}\left(\phi_{\mathbf{k}}\pm \frac{\pi}{2}\right)} \\
e^{\frac{i}{2}\left(\phi_{\mathbf{k}}\pm \frac{\pi}{2}\right)} \end{array} \right ).
\end{equation}
in Eq.~(\ref{surface1}), and $F_\pm(z) \sim e^{\lambda_\pm z}$. This leads to the following secular equations
\begin{eqnarray}
\lambda^2_\pm \mp \frac{\Delta_0}{\mu k_F} k^2_\perp \lambda_\pm -(k^2_\perp-k^2_F)=0.
\end{eqnarray}
The secular equations have the following solutions
\begin{eqnarray}
\lambda_{+,j}=\frac{\Delta_0 k^2_\perp}{2\mu k_F}+(-1)^j \sqrt{\left(1+\frac{\Delta^2_0k^2_\perp}{4\mu^2k^2_F}k^2_\perp \right)-k^2_F} \phantom{\hspace{7mm}}\\
\lambda_{-,j}=-\frac{\Delta_0 k^2_\perp}{2\mu k_F} +(-1)^j \sqrt{\left(1+\frac{\Delta^2_0k^2_\perp}{4\mu^2k^2_F}k^2_\perp \right)-k^2_F},\phantom{\hspace{3mm}}
\end{eqnarray}
where $j=1,2$. The boundary condition $\psi(z\to -\infty)=0$ can only be satisfied by $\psi_+(z) \sim e^{\lambda_{+,j}z}$, when $k_\perp < k_F$, and $k_\perp \neq 0$. In order to satisfy the boundary condition at $z=0$, we require $F_+(z)=A_1 \; \sum_j (-1)^j \; e^{\lambda_{+,j}z}$, and obtain
\begin{widetext}
\begin{eqnarray}
F_+(z)=\Theta(-z) \; \left[\frac{\Delta_0k^2_\perp}{\mu k_F}\frac{k^2_F-k^2_\perp}{k^2_F-\left(1+\frac{\Delta^2_0k^2_\perp}{4\mu^2k^2_F}k^2_\perp \right)}\right]^{\frac{1}{2}} \; \exp \left[\frac{\Delta_0k^2_\perp z}{2\mu k_F}\right]\; \sinh \left[z\sqrt{\left(1+\frac{\Delta^2_0k^2_\perp}{4\mu^2 k^2_F}k^2_\perp \right)-k^2_F}\right].
\end{eqnarray}
\end{widetext}

\section{Chiral Andreev bound states on the $\boldsymbol{(1,0,0)}$ and $\boldsymbol{(0,1,0)}$ surface}\label{Appendix2}

In this section we are considering a boundary in the $x$-direction, such that $x<0$ and $x>0$ regions are respectively occupied by the superconductor with $L_z=-2$ pairing and the vacuum. The two component spinor wavefunction now satisfies the differential equations
\begin{eqnarray}
&\bigg[&\phantom{\hspace{1mm}}\big(-\partial^2_x+k^2_{y}+k^2_z-k^2_F\big)\tau_3 -\frac{\Delta_0 k_z }{\mu k_F}\;\left(\partial^2_x +k^2_y \right)\tau_1 \nonumber \\
&\phantom{\hspace{1mm}}& - \frac{2i\Delta_0 k_z }{\mu k_F}k_y \partial_x\tau_2\bigg]\psi =2mE \psi,
 \end{eqnarray}
and the boundary conditions $\psi(0)=\psi(x\to -\infty)=0$. We notice that $-\partial^2_x$ and $\partial_x$ respectively appear with the matrices $\left(\tau_3+\frac{\Delta_0}{\mu}\tau_1\right)$ and $\tau_2$. The calculation can be simplified by rewriting
\begin{eqnarray}
\left(-\partial^2_x+k^2_{y}+k^2_z-k^2_F\right)\tau_3 -\frac{\Delta_0 k_z }{\mu k_F}\;\left(\partial^2_x +k^2_y \right)\tau_1 \nonumber \\
 =a \left(\tau_3+\frac{\Delta_0 k_z}{\mu k_F}\tau_1\right)+b \left(\tau_1- \frac{\Delta_0 k_z}{\mu k_F}\tau_3\right),
\end{eqnarray}
where
\begin{eqnarray}
&&a=-\partial^2_x+\frac{k^2_z+k^2_y(1-\Delta^2_0k^2_z/\mu^2 k^2_F)-k^2_F}{1+\Delta^2_0k^2_z/(\mu^2k^2_F)}, \\
&&b=\frac{\Delta_0 k_z}{\mu k_F}\frac{k^2_F-k^2_z-2k^2_y}{1+\Delta^2_0k^2_z/(\mu^2k^2_F)}.
\end{eqnarray}
Notice that $\mathcal{U}_3=\left(\tau_1- \frac{\Delta_0k_z}{\mu k_F}\tau_3\right)$ anticommutes with the Hamiltonian or the differential operator when $b=0$, and the zero energy states are thus protected by the spectral symmetry with respect to $\mathcal{U}_3$. The zero energy states are located at
\begin{equation}
k_y=\pm \frac{1}{\sqrt{2}}\sqrt{k^2_F-k^2_z},
\end{equation}
and describe two spin degenerate Majorana--Fermi arcs.

The index theorem mandates the existence of two independent zero energy solutions (without taking into account the spin degeneracy), which are eigenstates of $\mathcal{U}_3$ with eigenvalues $\mathrm{sgn}(k_y) \times \sqrt{1+\Delta^2_0/\mu^2}$. The differential equations are now solved by following the strategy of the previous subsection, and we obtain the following chiral dispersion relation
\begin{equation}
E(k_y, k_z \neq 0)=-\frac{\mathrm{sgn}(k_y)\Delta_0 \mu |k_z|}{\sqrt{\mu^2 k^2_F+ \Delta^2_0 k^2_z}} \left(1-\frac{k^2_z}{k^2_F}-\frac{2k^2_y}{k^2_F}\right). \label{energy}
\end{equation} and from $E(k_y, k_z \!\neq\! 0)=0$, we recover the Fermi arcs.


\begin{thebibliography}{10}

\bibitem{Wan} X. Wan, A. Turner, A. Vishwanath, and S. Y. Savrasov, Phys. Rev. B \textbf{83}, 205101 (2011).

\bibitem{Volovik} G. E. Volovik, \textit{Universe in a helium droplet}, Oxford University Press, (2003).

\bibitem{Beri} B. B\'{e}ri, Phys. Rev. B \textbf{81}, 134515 (2010).

\bibitem{Schnyder} A. P. Schnyder and S. Ryu, Phys. Rev. B \textbf{84}, 060504(R) (2011).

\bibitem{Kopnin} T. T. Heikkil\"{a}, N. B. Kopnin, and G. E. Volovik, JETP Lett. \textbf{94}, 233 (2011).

\bibitem{Schnyder_2008} A. P. Schnyder, S. Ryu, A. Furusaki, and A. W. W. Ludwig, Phys. Rev. B \textbf{78} 195125 (2008);

\bibitem{Kitaev_2009} A. Yu Kitaev AIP Conf. Proc. \textbf{1134} 22 (2009).

\bibitem{Ryu_2010} S. Ryu, A. Schnyder, A. Furusaki, A. W. W. Ludwig, New J. Phys. \textbf{12}, 065010 (2010).

\bibitem{GoswamiChakravarty} P. Goswami, and S. Chakravarty, Phys. Rev. Lett. \textbf{107}, 196803 (2011).

\bibitem{Burkov1} A. A. Burkov, and L. Balents, Phys. Rev. Lett. \textbf{107}, 127205 (2011).

\bibitem{Burkov2} A. A. Burkov, M. D. Hook, and L. Balents, Phys. Rev. B \textbf{84}, 235126 (2011).

\bibitem{Zyuzin1} A. A. Zyuzin, S. Wu, and A. A. Burkov, Phys. Rev. B \textbf{85}, 165110 (2012).

\bibitem{Cho} G. Y. Cho, arXiv:1110.1939 (2011).

\bibitem{Liu} C. X. Liu, P. Ye, X. L. Qi, Phys. Rev. B {\bf 87}, 235306; Erratum: Phys. Rev. B {\bf 92}, 119904 (2015).

\bibitem{Gong} M. Gong, S. Tewari, C. W. Zhang, Phys. Rev. Lett. \textbf{107}, 195303 (2011).

\bibitem{Sau} J. D. Sau, S. Tewari, Phys. Rev. B \textbf{86}, 104509 (2012).

\bibitem{Meng} T. Meng, L. Balents, Phys. Rev. B \textbf{86}, 054504 (2012).

\bibitem{Das} T. Das, Phys. Rev. B \textbf{88}, 035444 (2013).

\bibitem{GoswamiTewari1} P. Goswami, and S. Tewari, arXiv:1311.1506

\bibitem{GoswamiBalicas} P. Goswami and L. Balicas, arXiv:1312.3632

\bibitem{Bishop84} D. J. Bishop, C. M. Varma, B. Battlogg, and E. Bucher, Z. Fisk, and J. L. Smith, Phys. Rev. Lett. {\bf 53}, 1009 (1984). 

\bibitem{Kohori88} Y. Kohori, T. Kohara, H. Shibai, Y. Oda, Y. Kitaoka, and K. Asayama, J. Phys. Soc. Jpn. \textbf{57}, 395 (1988).  

\bibitem{Broholm90} C. Broholm, G. Aeppli, R. N. Kleiman, D. R. Harshman, D. J. Bishop, E. Bucher, D. Ll. Williams, E. J. Ansaldo, and R. H. Heffner, Phys. Rev. Lett. \textbf{65}, 2062 (1990).  

\bibitem{Suderow97} H. Suderow, J. P. Brison, A. Huxley, and J. Flouquet, J. Low Temp. Phys. \textbf{108}, 11 (1997).

\bibitem{Ramirez95} A. P. Ramirez, N. Stuchelli, and E. Bucher, Phys. Rev. Lett. \textbf{74}, 1218 (1995).

\bibitem{Joynt92} R. Joynt, J. Magn. Magn. Mater. {\bf 108}, 31 (1992).

\bibitem{Norman92} M. R. Norman, Physica C {\bf 194}, 203 (1992).

\bibitem{Sauls94} J. A. Sauls, Adv. Phys. {\bf 43}, 113 (1994).

\bibitem{Norman96} M. R. Norman and P. J. Hirschfeld, Phys. Rev. B {\bf 53}, 5706 (1996).

\bibitem{Graf00} M. J. Graf, S. Yip, and J. A. Sauls, Phys. Rev. B {\bf 62}, 14393 (2000).

\bibitem{Joynt02} R. Joynt and L. Taillefer, Rev. Mod. Phys. {\bf 74}, 235 (2002).

\bibitem{Tou96} H. Tou, Y. Kitaoka, K. Asayama, N. Kimura, Y. ?nuki, E. Yamamoto, and K. Maezawa, Phys. Rev. Lett. \textbf{77}, 1374 (1996). 

\bibitem{Tou98} H. Tou, Y. Kitaoka, K. Ishida, K. Asayama, N. Kimura, Y. ?nuki, E. Yamamoto, Y. Haga, and K. Maezawa, Phys. Rev. Lett. \textbf{80}, 3129 (1998). 

\bibitem{Strand09} J. D. Strand, D. J. Van Harlingen, J. B. Kycia, and W. P. Halperin, Phys. Rev. Lett. \textbf{103}, 197002 (2009).

\bibitem{Strand10} J. D. Strand, D. J. Bahr, D. J. Van Harlingen, J. P. Davis, W. J. Gannon, W. P. Halperin, Science \textbf{328}, 1368 (2010).  

\bibitem{footnote1}We note that such a spin-triplet  $E_{2u}$ order parameter in the $B$-phase of UPt$_3$ seemingly violates Blount's theorem~\cite{Blount85}, which states that the line nodes are unstable in spin-triplet supercodunctors. However, it has been shown in a recent study by Kobayashi and co-workers~\cite{Kobayashi} that Blount's proof implicitly assumes time-reversal symmetry, whereas it is spontaneously broken in the $B$-phase (see below). As a result, the $B$-phase of UPt$_3$ falls into the so-called $M^{+}$ symmetry class, which Kobayashi \emph{et al.} proved to be a counter-example to Blount's theorem, which does permit stable line nodes~\cite{Kobayashi}.


\bibitem{Blount85} E. I. Blount, Phys. Rev. B \textbf{32}, 2935 (1985).

\bibitem{Kobayashi} S. Kobayashi, K. Shiozaki, Y. Tanaka, and M. Sato, Phys. Rev. B {\bf 90}, 024516 (2014).

\bibitem{Machida12} Y. Machida, A. Itoh, Y. So, K. Izawa, Y. Haga, E. Yamamoto, N. Kimura, Y. Onuki, Y. Tsutsumi, and K. Machida, Phys. Rev. Lett. {\bf 108}, 157002 (2012). 

\bibitem{Tsutsumi1}Y. Tsutsumi, K. Machida, T. Ohmi, and M. Ozaki, J.~Phys.~Soc.~Jpn. {\bf 81}, 074717 (2012).

\bibitem{Tsutsumi2}Y. Tsustsumi,  M. Ishikawa, T. Kawakami, T. Mizushima, M. Sato, M. Ichioka, K. Machida, J. Phys. Soc. Jpn. {\bf 82}, 113707 (2013). 

\bibitem{Izawa2014} K. Izawa, Y. Machida, A. Itoh, Y. So, K. Ota, Y. Haga, E. Yamamoto, N. Kimura, Y. Onuki, Y. Tsutsumi, and K. Machida, J. Phys. Soc. Jpn {\bf 83}, 061013 (2014). 

\bibitem{Luke93} G. M. Luke, A. Keren, L. P. Le, W. D. Wu, Y. J. Uemura, D. A. Bonn, L. Taillefer, and J. D. Garrett, Phys. Rev. Lett. {\bf 71}, 1466 (1993).

\bibitem{footnote2} Although this result was not reproduced in a later $\mu$SR experiment \cite{Reotier95}.

\bibitem{Reotier95} P. D. Reotier, A. Huxley, A. Yaouanc, J. Flouquet, P. Bonville, P. Imbert, P. Pari, P.C.M. Gubbens, and A.M. Mulders, Phys. Lett. A {\bf 205}, 239 (1995). 

\bibitem{Kapitulnik} E. R. Schemm, W. J. Gannon, C. M. Wishne, W. P. Halperin, and A. Kapitulnik, Science 345, 190-193 (2014).




\bibitem{footnote3} There is no symmetry requirement for the intrinsic Kerr effect to vanish in UPt$_3$, contrary to the argument (see e.g. Ref.~\onlinecite{Mineev12}) that optical conductivity must be zero in a translationally-invariant superconductor. The optical conductivity only vanishes in the presence of Galilean invariance, which in a real material can be broken by the interband matrix elements ~\cite{Gradhand13, Mineev13}.

\bibitem{Mineev12} V. P. Mineev, J. Phys. Soc. Jpn. {\bf 81}, 093703 (2012).

\bibitem{Gradhand13}  M. Gradhand, K. I. Wysokinski, J. F. Annett, and B. L. Gy\"orffy, Phys. Rev. B {\bf 88}, 094504 (2013).

\bibitem{Mineev13} V. P. Mineev, Phys. Rev. B {\bf 89}, 134519 (2014).

\bibitem{Tanaka1} M. Sato, Y. Tanaka, K. Yada, and T. Yokoyama, Phys. Rev. B {\bf 83}, 224511 (2011).

\bibitem{Tanaka2} Y. Tanaka, M. Sato, and N. Nagaosa, J. Phys. Soc. Jpn. {\bf 81}, 011013 (2012).

\bibitem{Moler1} J. R. Kirtley, C. Kallin, C. W. Hicks, E.-A. Kim, Y. Liu, K. A. Moler, Y. Maeno, and K. D. Nelson, Phys. Rev. B \textbf{76}, 014526 (2007).

\bibitem{GoswamiTewari2} P. Goswami, and S. Tewari, Phys. Rev. B \textbf{88}, 245107 (2013).

\bibitem{Lutchyn} R. M. Lutchyn, P. Nagornykh, and V. M. Yakovenko, Phys. Rev. B \textbf{77}, 144516 (2008).

\bibitem{Kallin} R. Roy, and C. Kallin, Phys. Rev. B \textbf{77}, 174513 (2008).

\bibitem{Read} N. Read and D. Green, Phys. Rev. B \textbf{61}, 10267 (2000).

\bibitem{Vafek} O. Vafek, A. Melikyan, and Z. Tesanovic, Phys. Rev. B \textbf{64}, 224508 (2001).

\bibitem{McMullan} G. J. McMullan, P. M. C. Rourke, M. R. Norman, A. D. Huxley, N. Doiron-Leyraud, J. Flouquet, G. G. Lonzarich, A. McCollam, and S. R. Julian, New J. Phys. {\bf 10}, 053029 (2008).

\bibitem{Checkelsky12} J. G. Checkelsky, R. Thomale, L. Li, G. F. Chen, J. L. Luo, N. L. Wang, and N. P. Ong, Phys. Rev. B {\bf 86}, 180502(R) (2012).

\bibitem{Hirschfeld1}P.J. Hirschfeld, P. W\"olfle and D. Einzel, Phys. Rev. B \textbf{37} 83, (1988).

\bibitem{Ott}H. R. Ott, E. Felder, C. Bruder, and T. M. Rice, Europhys. Lett. 3,  1123(1987).

\end{thebibliography}
\end{document}